\pdfoutput=1
\documentclass[conference]{IEEEtran}
\usepackage{cite}
\ifCLASSINFOpdf
  \usepackage[pdftex]{graphicx}
  \usepackage{subfig}

\else
\fi
\usepackage[cmex10]{amsmath}
\makeatletter
\let\@ORGmakecaption\@makecaption
\long\def\@makecaption#1#2{\@ORGmakecaption{#1}{#2}\vskip\belowcaptionskip\relax}
\makeatother
\usepackage{listings}
\usepackage[labelsep=period]{caption}
\begin{document}
%
\title{Mira: A Framework for Static Performance Analysis}


\author{\IEEEauthorblockN{Kewen Meng, Boyana Norris}
\IEEEauthorblockA{Department of Computer and Information Science\\
University of Oregon\\
Eugene, Oregon\\
\{kewen, norris\}@cs.uoregon.edu}
}


%


\maketitle

\begin{abstract}
    The performance model of an application can provide understanding about its runtime behavior on particular hardware. Such information can be analyzed by developers for performance tuning. However, model building and analyzing is frequently ignored during software development until performance problems arise because they require significant expertise and can involve many time-consuming application runs. In this paper, we propose a fast, accurate, flexible and user-friendly tool, Mira, for generating performance models by applying static program analysis, targeting scientific applications running on supercomputers. We parse both the source code and binary to estimate performance attributes with better accuracy than considering just source or just binary code. Because our analysis is static, the target program does not need to be executed on the target architecture, which enables users to perform analysis on available machines instead of conducting expensive experiments on potentially expensive resources. Moreover, statically generated models enable performance prediction on non-existent or unavailable architectures. In addition to flexibility, because model generation time is significantly reduced compared to dynamic analysis approaches, our method is suitable for rapid application performance analysis and improvement. We present several scientific application validation results to demonstrate the current capabilities of our approach on small benchmarks and a mini application.
\end{abstract}


%
\IEEEpeerreviewmaketitle

\section{Introduction}
Understanding application and system performance plays a critical role in high-performance software and architecture design. Developers require a thorough insight of the application and system to aid their development and improvement. Performance modeling provides software developers with necessary information about bottlenecks and can guide users in identifying potential optimization opportunities. 

As the development of new hardware and architectures progresses, the computing capability of high performance computing (HPC) systems continues to increase dramatically. However along with the rise in computing capability, it is also true that many applications cannot use the full available computing potential, which wastes a considerable amount of computing power. The inability to fully utilize available computing resources or specific advantages of architectures during application development partially accounts for this waste. Hence, it is important to be able to understand and model program behavior in order to gain more information about its bottlenecks and performance potential. Analyzing the instruction mixes of programs at function or loop granularity can provide insight on CPU and memory characteristics, which can be used for further optimization of a program. 

In this paper, we introduce a new approach for analyzing and modeling programs using primarily static analysis techniques combining both source and binary program information. Our tool, Mira, generates parameterized performance models that can be used to estimate instruction mixes at different granularity (from function to statement level) for different inputs and architectural features without requiring execution of the application.

\begin{figure*}[hbt]
  \centering
    \includegraphics[width=0.8\textwidth]{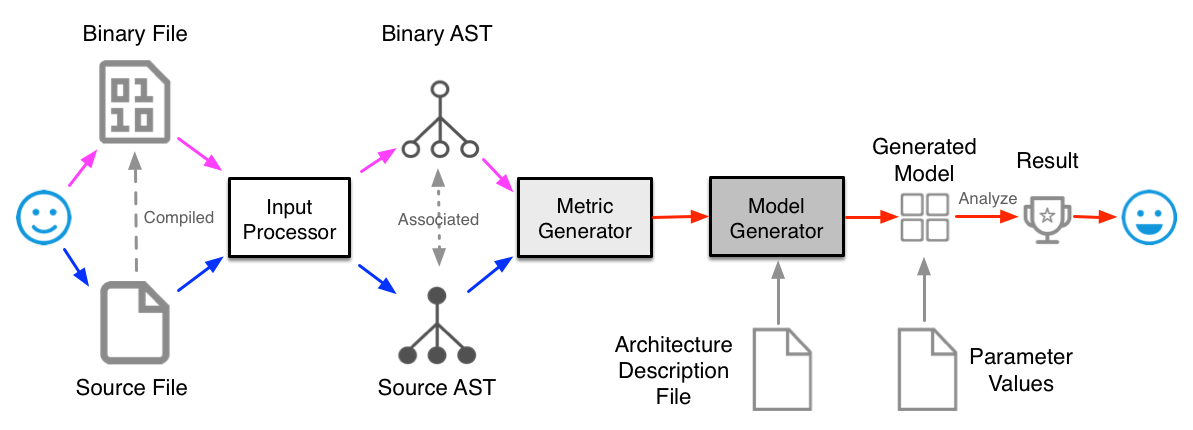}
  \caption{Workflow of Mira for generation of performance model and analysis.}
  \label{fig:workflow}
\end{figure*}

Current program performance analysis tools can be categorized into two types: static and dynamic. Dynamic (runtime) analysis is performed through execution of the target program and measurement of metrics of interest, e.g., time or hardware performance counters. By contrast, static analysis operates on the source or binary code without actually executing it. PBound~\cite{PBound} is an example static analysis tool for automatically modeling program performance based on source code analysis of C applications. Because PBound considers only the source code, it cannot capture compiler optimizations and hence produces less accurate estimates of performance metrics. We discuss other examples of these approaches in more detail in Sections~\ref{sec:background} and ~\ref{sec:related}. 

While some past research efforts mix static and dynamic analysis to create a performance model, relatively little effort has been put into pure static performance analysis and increasing the accuracy of static analysis. Our approach starts from object code because the code transformations performed by optimizing compilers would cause non-negligible effects on the analysis accuracy. In addition, object code is language-independent and more directly reflects runtime behavior. Although object code could provide instruction-level information, it still fails to offer some critical factors for understanding the target program. For instance, it is difficult or impossible to obtain detailed information about high-level code structures (user-defined types, classes, loops) from just the object code. Therefore, source code is also analyzed in our project to supplement complementary high-level information. 

By combining source code and object code representations, we are able to obtain a more precise description of the program and its possible behavior when running on a particular architecture, which results in improved modeling accuracy. The output of our tool can be used to rapidly explore program behavior for different inputs without requiring actual application execution. In addition, because the analysis is parameterized with respect to the architecture, Mira provides users valuable insight of how programs may run on particular architectures without requiring access to the actual hardware. Furthermore, the output of Mira can also be applied to create performance models to further analyze or optimize performance, for example Roofline arithmetic intensity estimates~\cite{williams2009roofline}.

This paper is organized as follows: Section~\ref{sec:background} briefly describes the ROSE compiler framework, the polyhedral model for loop analysis, and the background of performance measurement and analysis tools. In Sections~\ref{sec:approach}, we discuss the details of our methodology and the implementation. Section~\ref{sec:result} evaluates the accuracy of the generated models on several benchmark codes. In Section~\ref{sec:related}, we introduce related work about static performance modeling. Section~\ref{sec:conclusion} concludes with a summary and future work discussion.

\section{Background}\label{sec:background}
\subsection{ROSE Compiler Framework}

ROSE~\cite{rose} is an open-source compiler framework developed at Lawrence Livermore National Laboratory (LLNL). It supports the development of source-to-source program transformation and analysis tools for large-scale Fortran, C, C++, OpenMP and UPC (Unified Parallel C) applications. ROSE uses the EDG (Edison Design Group) parser and OPF (Open Fortran Parser) as the front-ends to parse C/C++ and Fortran. The front-end produces ROSE intermediate representation (IR) that is then converted into an abstract syntax tree (AST). It provides users a number of APIs for program analysis and transformation, such as call graph analysis, control flow analysis, and data flow analysis. The wealth of available analyses makes ROSE an ideal tools both for experienced compiler researchers and tool developers with minimal background to build custom tools for static analysis, program optimization, and performance analysis.

\subsection{Polyhedral Model}
We rely on the polyhedral model to characterize the iteration spaces of certain types of loops. The polyhedral model is an intuitive algebraic representation that treats each loop iteration as lattice point inside the polyhedral space produced by loop bounds and conditions. Nested loops can be translated into a polyhedral representation if and only if they have affine bounds and conditional expressions, and the polyhedral space generated from them is a convex set. Moreover, the polyhedral model can be used to generate generic representation depending on loop parameters to describe the loop iteration domain. In addition to program transformation~\cite{pouchet.11.popl}, the polyhedral model is broadly used for automating optimization and parallelization in compilers (e.g., GLooG~\cite{Bas04b}) and other tools~\cite{griebl2004automatic,bondhugula2008practical,grosser2012polly}.

\subsection{Performance Measurement and Analysis Tools}
Performance tools are capable of gathering performance metrics either dynamically (instrumentation, sampling) or statically. PAPI~\cite{mucci1999papi} is used to access hardware performance counters through both high- and low-level interfaces, which are typically used through manual or automated instrumentation of application source code. The high-level interface supports simple measurement and event-related functionality such as start, stop or read, whereas the low-level interface is designed to deal with more complicated needs. The Tuning and Analysis Utilities (TAU)~\cite{shende2006tau} is another state-of-the-art performance tool that uses PAPI as the low-level interface to gather hardware counter data. TAU is able to monitor and collect performance metrics by instrumentation or event-based sampling. In addition, TAU also has a performance database for data storage and analysis and visualization components, ParaProf. There are several similar performance tools including HPCToolkit~\cite{adhianto2010hpctoolkit}, Scalasca~\cite{geimer2010scalasca}, MIAMI~\cite{marin2014miami}, gprof~\cite{graham1982gprof}, Byfl~\cite{pakin2013hardware}, which can also be used to analyze application or systems performance through runtime measurements.

\section{Approach}\label{sec:approach}
Mira is built on top of ROSE compiler framework, which provides several useful APIs as front-end for parsing the source file and disassembling the ELF file. Mira is implemented in C++ and is able to process C/C++ source code as input. Figure~\ref{fig:workflow} illustrates the entire workflow of Mira for performance model generation and analysis, which comprises three major parts:
\begin{itemize}
  \item \textbf{Input Processor} - Input parsing and disassembling
  \item \textbf{Metric Generator} - AST traversal and metric generation
  \item \textbf{Model Generator} - Model generation in Python
\end{itemize}

\subsection{Processing Input Files}
\subsubsection{Source code and binary representations}
The \textbf{Input Processor} is the front-end of Mira, and its primary goal is to process source code and ELF object file inputs and build the corresponding ASTs (Abstract Syntax Trees). Mira analyzes these ASTs to locate critical structures such as function bodies, loops, and branches. Furthermore, because the source AST also preserves high-level source information, such as variable names, types, the order of statements and the right/left hand side of assignment, Mira incorporates this high-level information into the generated model. For instance, one can query all information about the static control part (SCoP) of a loop, including loop initialization, loop condition, and increment (these are not explicit in the binary code). In addition, because variable names are preserved, it makes the identification of loop indexes much easier and processing of the variables inside the loop more accurate. 

\begin{figure}
  \centering
    \includegraphics[width=0.4\textwidth]{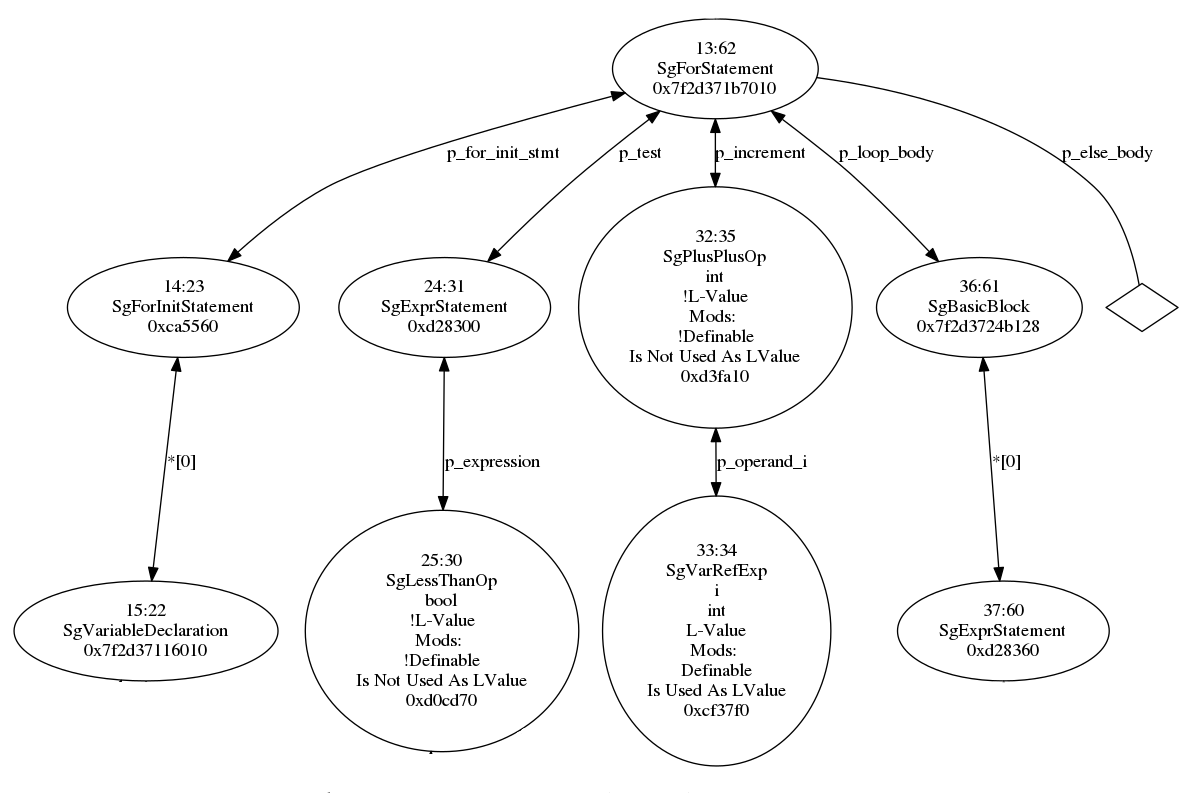}
  \caption{Loop structure from a C++ source code AST fragment (ROSE-generated dot graph output).}
  \label{fig:partial_loop}
\end{figure}

\subsubsection{Bridge between source and binary}
The AST is the output of the frond-end part of Mira. After processing the inputs, two ASTs are generated separately from the source and compiled binary codes representing the structures of the two inputs. Mira is designed to use information retrieved from these trees to improve the accuracy of the generated models. Therefore, it is necessary to build connections between the two ASTs so that for a structure in source it is able to instantly locate corresponding nodes in the binary one.

\begin{figure}
  \centering
    \includegraphics[width=0.4\textwidth]{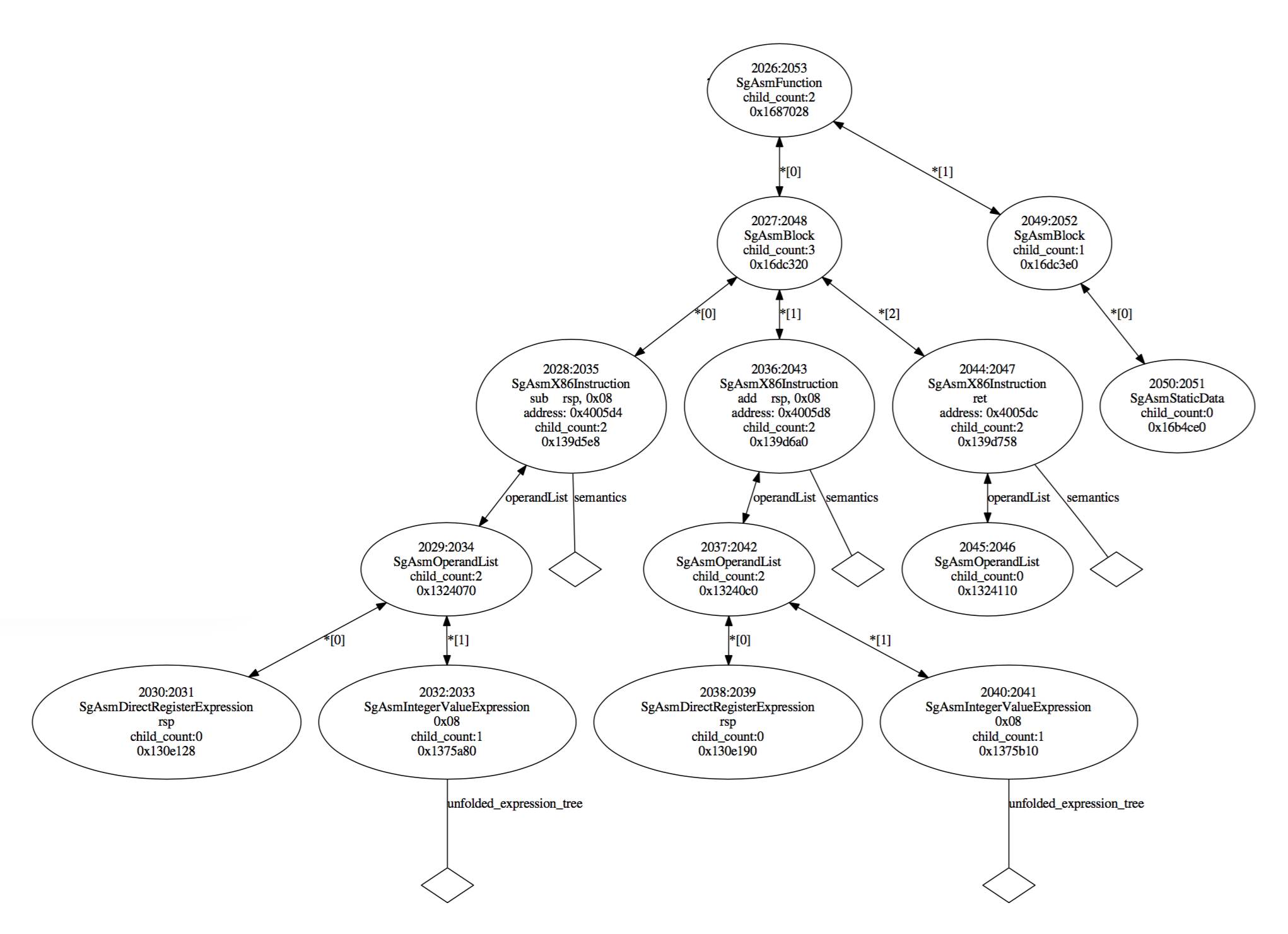}
  \caption{Partial binary AST (ROSE-generated dot graph output).}
  \label{fig:binAST}
\end{figure}

Although both ASTs are representations of the inputs, they have totally different shapes, node organizations and meaning of nodes. A partial binary AST (representing a function) is shown in Figure~\ref{fig:binAST}. Each node of the binary AST describes the syntax element of assembly code, such as \textit{SgAsmFunction}, \textit{SgAsmX86Instruction}. As shown in Figure~\ref{fig:binAST}, a function in the binary AST is composed of multiple instructions, while in the source AST, a functions is composed of statements. Hence, one source AST node typically corresponds to several nodes in the binary AST, which complicates the building of connections between them.

Because the differences between the two AST structures make it difficult to connect source to binary, an alternate way is needed to make the connection between ASTs more precise. Inspired by debuggers, line numbers are used in our tool as the bridge to associate source to binary. When we are debugging a program, the debugger knows exactly the source line and column of the error location. By using the \textit{-g} option during program compilation, the compiler will insert debug-related information into the object file for future reference. Most compilers and debuggers use DWARF (debugging with attributed record format) as the debugging file format to organize the information for source-level debugging. DWARF categorizes data into several sections, such as \textit{.debug\_info}, \textit{.debug\_frame}, etc. The  \textit{.debug\_line} section stores the line number information. 

The line number debugging information allows us to decode the specific DWARF section to map the line number to the corresponding instruction address. Because line number information in the source AST is already preserved in each node, unlike the binary AST, it can be retrieved directly. After line numbers are obtained from both source and binary, connections are built in each direction between the two ASTs. As mentioned in the previous section, a source AST node normally links to several binary AST nodes due to the different meaning of nodes. Specifically, a statement contains several instructions, but an instruction only has one connected source location. Once the node in the binary AST is associated to the source location, further analysis can be performed. For instance, it is possible to narrow the analysis to a small scope and collect data such as the instruction count and type in a particular code fragment, such as function body, loop body, and even a single statement. 

\begin{table*}[t]
\centering
\caption{Loop coverage in high-performance applications}
\label{tab:1}
\begin{tabular}{|l|l|l|l|l|}
\hline
Application & Number of loops & Number of statements & Statements in loops & Percentage \\ \hline\hline
applu       & 19              & 757                  & 633                 & 84\%       \\ \hline
apsi        & 80              & 2192                 & 1839                & 84\%       \\ \hline
mdg         & 17              & 530                  & 464                 & 88\%       \\ \hline
lucas       & 4               & 2070                 & 2050                & 99\%       \\ \hline
mgrid       & 12              & 369                  & 369                 & 100\%      \\ \hline
quake       & 20              & 639                  & 489                 & 77\%       \\ \hline
swim        & 6               & 123                  & 123                 & 100\%      \\ \hline
adm         & 80              & 2260                 & 1899                & 84\%       \\ \hline
dyfesm      & 75              & 1497                 & 1280                & 86\%       \\ \hline
mg3d        & 39              & 1442                 & 1242                & 86\%       \\ \hline
\end{tabular}
\end{table*}

\subsection{Generating metrics}
The \textbf{Metric Generator} is an important part of the entire framework, which has significant impact on the accuracy of the generated model. It receives the ASTs as inputs from the Input Processor to produce metrics for model generation. An AST traversal is needed to collect and propagate necessary information about the specific structures in the program for appropriate organization of the program representation to precisely guide model generation. During the AST traversal, additional information is attached to the particular tree node as a supplement used for analysis and modeling. For example, if it is too long, one statement is probably located in several lines. In this case, all the line numbers are collected together and stored as extra information attached to the statement node. 

To best model the program, the metric generator traverses the source AST twice: first bottom-up and then top-down. The upward traversal propagates detailed information about specific structures up to the head node of the sub-tree. For instance, as shown in Figure~\ref{fig:partial_loop}, \textit{SgForStatement} is the head node for the loop sub-tree; however, this node itself does not store any information about the loop. Instead, the loop information such as loop initialization, loop condition and step are stored in \textit{SgForInitStatement}, \textit{SgExprStatement} and \textit{SgPlusPlusOp} separately as child nodes. In this case, the bottom-up traversal recursively collects information from leaves to root and organizes it as extra data attached to the head node for the loop. The attached information will serve as context in modeling. 

After bottom-up traversal, top-down traversal is applied to the AST. Because information about sub-tree structure has been collected and attached, the downward traversal primarily focuses on the head node of sub-tree and those of interest, for example the \textit{loop} head node, \textit{if} head node, \textit{function} head node, and \textit{assignment} node, etc. Moreover, the top-down traversal must pass down necessary information from parent to child node in order to model complicated structures correctly. For example, in nested loop and branch inside loop the inner structure requires the information from the parent node as the outer context to model itself, otherwise these complicated structures can not be correctly handled. Also, instruction information from ELF AST is connected and associated to correspond structures in top-down traversal. 

\subsection{Generating Models}\label{sec:model}
The \textbf{Model Generator} is built on the Metric Generator, which consumes the intermediate analysis result of the metric generator and generates an easy-to-use model. To achieve the flexibility, the generated model is coded in Python so that the result of the model can be directly applied to various scientific libraries for further analysis and visualization. In some cases, the model is in ready-to-execute condition for which users are able to run it directly without providing any input. However, users are required to feed extra input in order to run the model when the model contains parametric expressions. The parametric expression exists in the model because our static analysis is not able to handle some cases. For example, when user input is expected in the source code or the value of a variable comes from the returning of a call, the variables are preserved in the model as parameters that will be specified by the users before running the model.

%


\begin{figure*}
\centering

\subfloat[Polyhedral representation for double-nested loop]{
	\includegraphics[width=0.3\textwidth]{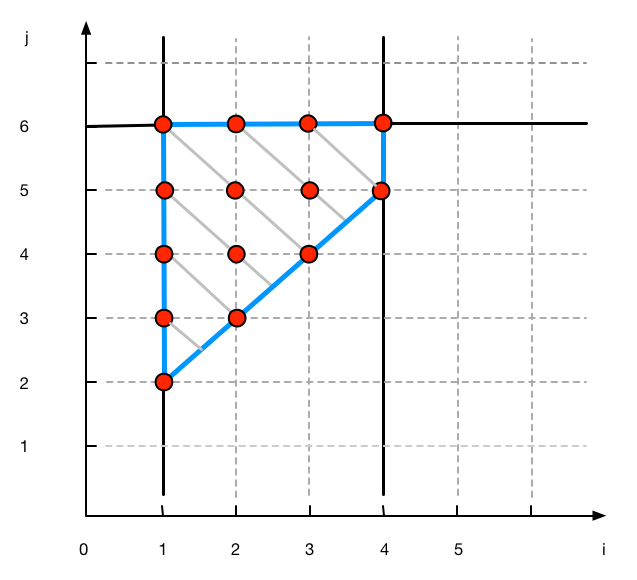}
	\label{loop:a}
}
\subfloat[Polyhedral representation with \textit{if} constraint]{
	\includegraphics[width=0.3\textwidth]{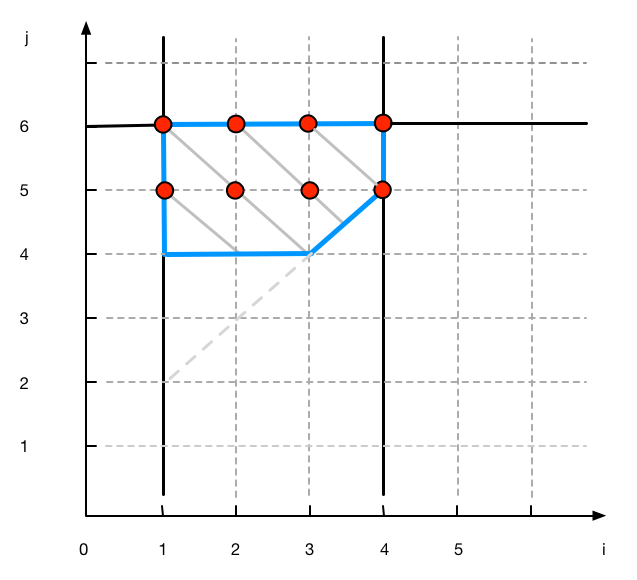}
	\label{loop:b}
}	
\hfill
\subfloat[\textit{if} constraint causing holes in the polyhedral area]{
	\includegraphics[width=0.3\textwidth]{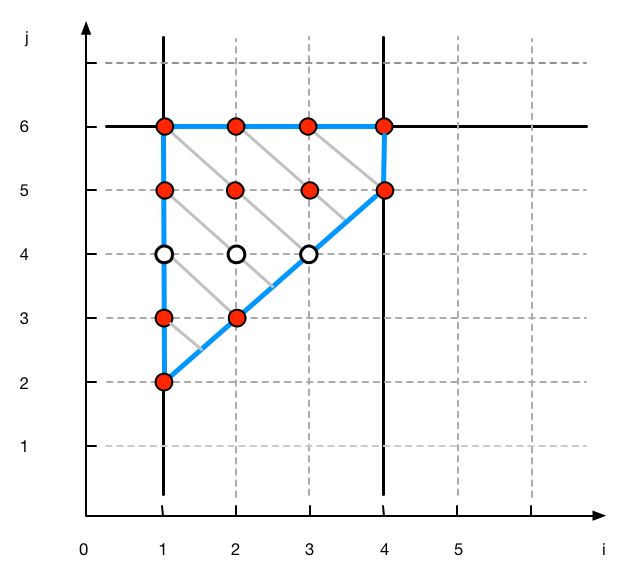}
	\label{loop:c}
}
\subfloat[Exceptions in polyhedral modeling]{
	\includegraphics[width=0.3\textwidth]{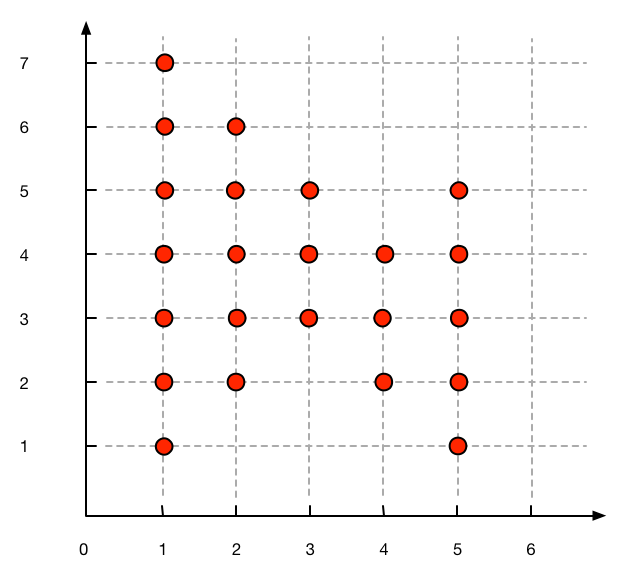}
	\label{loop:d}
}
\caption{Polyhedral model for a double-nested loop.}
\label{fig:loop}
\end{figure*}

\subsubsection{Loop Modeling}
Loops are common in HPC codes and are typically at the heart of the most time-consuming computations. A loop executes a block of code repeatedly until certain conditions are satisfied. Bastoul et al.~\cite{bastoul2003putting} surveyed multiple high-performance applications and summarized the results in Table~\ref{tab:1}. The first column shows the number of loops contained in the application. The second column lists the total number of statements in the applications and the third column counts the number of statements covered by loop scope. The ratio of in-loop statements to the total number of statements are calculated in the last column. In the data shown in the table, the lowest loop coverage is 77\% for \textit{quake} and the coverage rates for the rest of applications are above 80\%. This survey data also indicates that the in-loop statements make up a large majority portion of the total statements in the selected high-performance applications. 

\begin{lstlisting}[caption={Basic loop}, frame=bt, belowcaptionskip=0.8em, label={list:1}]
    for (i = 0; i < 10; i++)
    {
        statements;
    }
\end{lstlisting}

\subsubsection{Using the Polyhedral Model}
Whether loops can be precisely described and modeled has a direct impact on the accuracy of the generated model because the information about loops will be provided as context for further in-loop analysis. The term "loop modeling" refers to analysis of the static control parts (SCoP) of a loop to obtain the information about the loop iteration domain, which includes understanding of the initialization, termination condition and step. Unlike dynamic analysis tools which may collect runtime information during execution, our approach runs statically so the loop modeling primarily relies on SCoP parsing and analyzing. Usually to model a loop, it is necessary to take several factors into consideration, such as depth, data dependencies, bounds, etc. Listing~\ref{list:1} shows a basic loop structure, the SCoP is complete and simple without any unknown variable. For this case, it is possible to retrieve the initial value, upper bound and steps from the AST, then calculate the number of iterations. The iteration count is used as context when analyzing the loop body. For example, if corresponding instructions are obtained the from binary AST for the statements in Listing~\ref{list:1}, the actual count of these instructions is expected to be multiplied by the iteration count to describe the real situation during runtime. 

\begin{lstlisting}[caption={Double-nested loop}, frame=bt, belowcaptionskip=0.8em, label={list:2}]  
   for(i = 1; i <= 4; i++)
    for(j = i + 1; j <= 6; j++)
    {
	statements;
    }
\end{lstlisting}

However, loops in real application are more complicated, which requires our framework to handle as many scenarios as possible. Therefore, the challenge for modeling the loop is to create a general method for various cases. To address this problem, we use the polyhedral model in Mira to accurately model the loop. The polyhedral model is capable of handling an N-dimensional nested loop and represents the iteration domain in an N-dimensional polyhedral space. For some cases, the index of inner loop has a dependency with the outer loop index. As shown in Listing~\ref{list:2}, the initial value of the inner index \textit{j} is based on the value of the outer index \textit{i}. For this case, it is possible to derive a formula as the mathematical model to represent this loop, but it would be difficult and time-consuming. Most importantly, it is not general; the derived formula may not fit for other scenarios. To use the polyhedral model for this loop, the first step is to represent loop bounds in affine functions. The bounds for the outer and inner loop are $1 \leq i \leq 4$ and $i+1 \leq j \leq 6$, which can be written as two equations separately: 
\begin{align*}
\begin{bmatrix} 1 & 0 \\ -1 & 0 \end{bmatrix} \times \left[ \begin{array}{c} i \\ j \end{array} \right] + \left[ \begin{array}{c} -1 \\ 4 \end{array} \right] \geq 0 
\end{align*}

\begin{align*}
\begin{bmatrix} -1 & 1 \\ 0 & -1 \end{bmatrix} \times \left[ \begin{array}{c} i \\ j \end{array} \right] + \left[ \begin{array}{c} -1 \\ 6 \end{array} \right] \geq 0 
\end{align*}

In Figure~\subref*{loop:a}, the two-dimensional polyhedral area presenting the loop iteration domain is created based on the two linear equations. Each dot in the figure represents a pair of loop indexes (\textit{i}, \textit{j}), which corresponds to one iteration of the loop. Therefore, by counting the integers in the polyhedral space, we are able to parse the loop iteration domain and obtain the iteration times. For loops with more complicated SCoP, such as the ones contain variables instead of concrete numerical values, the polyhedral model is also applicable. When modeling loops with unknown variables, Mira uses the polyhedral model to generate a parametric expression representing the iteration domain, which can be changed by specifying different values to the input. Mira maintains the generated parametric expressions and uses as context in the following analysis. In addition, the unknown variables in loop SCoP are preserved as parameters until the parametric model is generated. With the parametric model, it is not necessary for the users to re-generate the model for different values of the parameters. Instead, they just have to adjust the inputs for the model and run the Python code to produce a concrete value. 
\begin{lstlisting}[caption={Exception in polyhedral modeling}, frame=bt, belowcaptionskip=0.8em, label={list:3}]  
   for(i = 1; i <= 5; i++)
    for(j = min(6 - i, 3); 
           j <= max(8 - i, i); j++)
    {
     statements;
    }
\end{lstlisting}
There are exceptions that the polyhedral model cannot handle. For the code snippet in Listing~\ref{list:3}, the SCoP of the loop forms a non-convex set (Figure~\subref*{loop:d}) which is not handled by the polyhedral model. Another problem in this code is that the loop initial value and loop bound depend on the return values of function calls. For static analysis to track and obtain the such values, more complex interprocedural analysis is required, which is planned as part of our future work.

\begin{lstlisting}[caption={Loop with \textit{if} constraint}, frame=bt, belowcaptionskip=0.8em, label={list:4}]  
   for(i = 1; i <= 4; i++)
    for(j = i + 1; j <= 6; j++)
    {
     if(j > 4)
     {
      statements;
     }    
    }
\end{lstlisting}

\subsubsection{Branches}
In addition to loops, branch statements are also common structures. In scientific applications, branch statements are frequently used to verify the intermediate output during the computing. Branch statements can be handled by the information retrieved from the AST. However, it complicates the analysis when the branch statements reside in a loop. In Listing~\ref{list:4}, the \textit{if} constraint $j > 4 $ is introduced into the previous code snippet. The number of execution times of the statement inside the \textit{if} depends on the branch condition. In our analysis, the polyhedral model of a loop is kept and passed down to the inner scope. Thus the \textit{if} node has the information of its outer scope.  Because the loop conditions combined with branch conditions form a polyhedral space as well, shown in Figure~\subref*{loop:b}, the polyhedral representation is still able to model this scenario by adding the branch constraint and regenerate a new polyhedral model for the \textit{if} node. Comparing Figure~\subref*{loop:b} with Figure~\subref*{loop:a}, it is obvious that the iteration domain becomes smaller and the number of integers decreases after introducing the constraint, which indicates the execution times of statements in the branch is limited by the \textit{if} condition.

\begin{lstlisting}[caption={\textit{if} constraint breaks polyhedral space}, frame=bt, belowcaptionskip=0.8em, label={list:5}]  
   for(i = 1; i <= 4; i++)
    for(j = i + 1; j <= 6; j++)
    {
     if(j % 4 != 0)
     {
      statements;
     }    
    }
\end{lstlisting}

\begin{figure*}
\centering

\subfloat[Source code]{
	\includegraphics[width=0.3\textwidth]{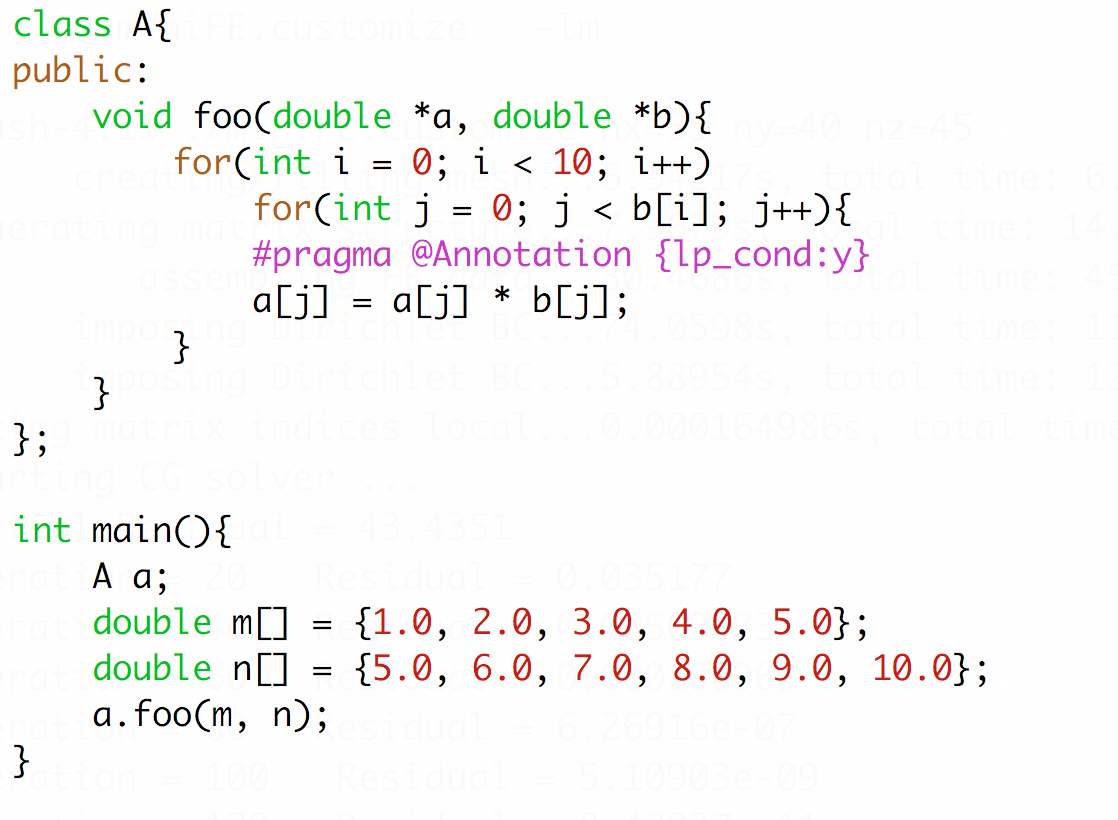}
	\label{model:a}
}
\subfloat[Generated \textit{foo} function]{
	\includegraphics[width=0.3\textwidth]{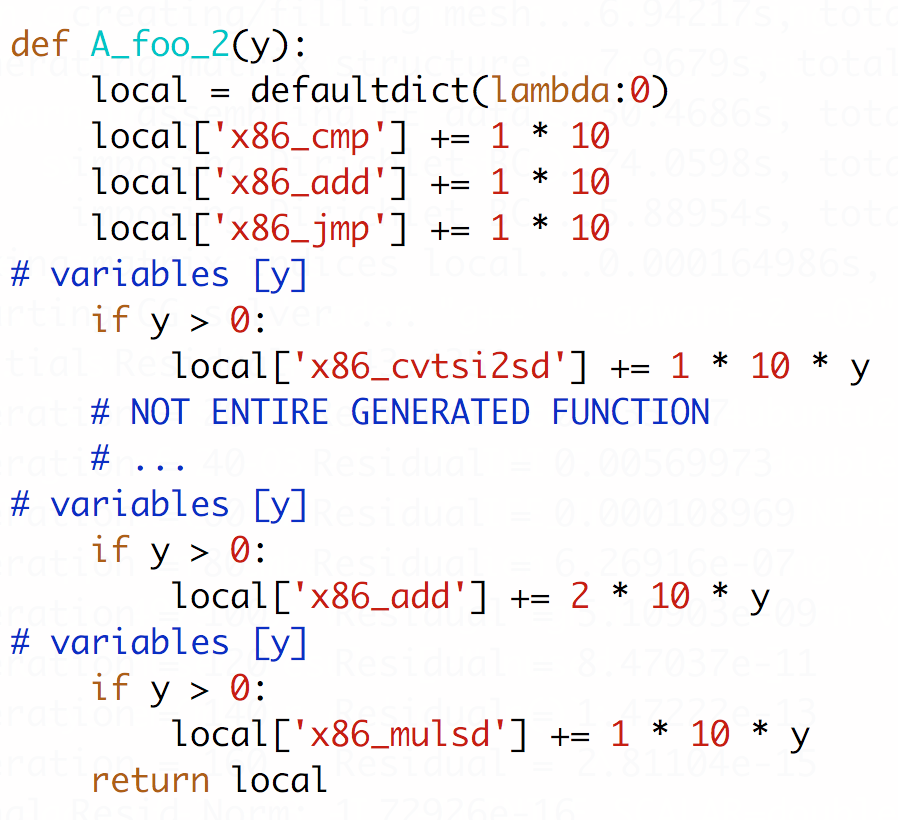}
	\label{model:b}
}
\subfloat[Generated \textit{main} function]{
	\includegraphics[width=0.3\textwidth]{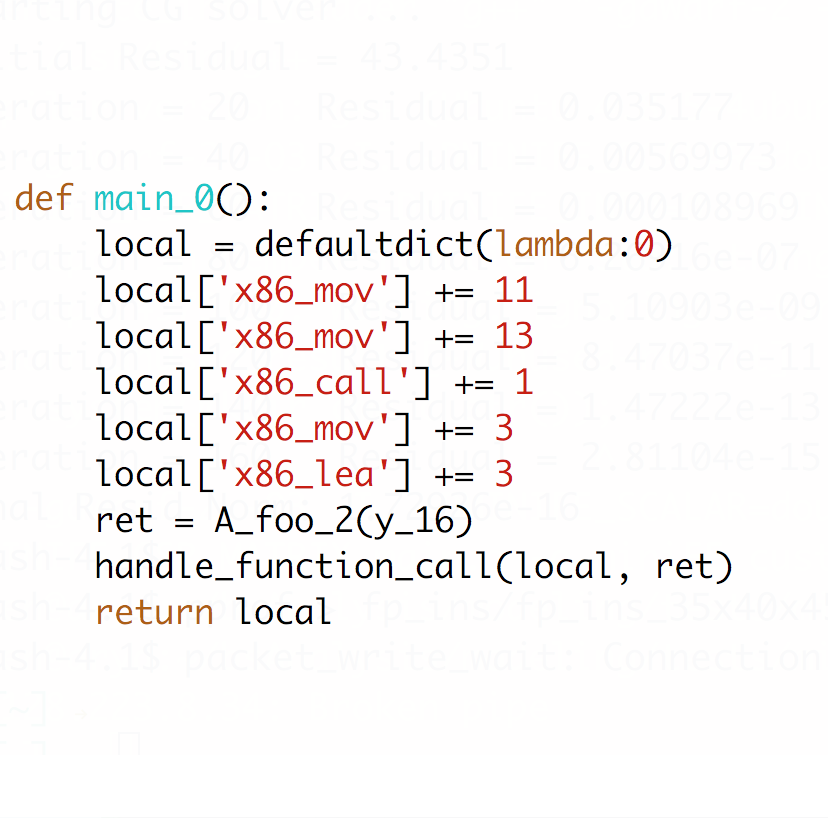}
	\label{model:c}
}	

\caption{Statically generated model: (a) Original source code; (b) Python model code excerpt for the \textit{foo} function; and (c) Python model code for the main program. }
\label{fig:model}
\end{figure*}

However, some branch constraints might break the definition of a convex set that the polyhedral model is not applicable. For the code in Listing~\ref{list:5}, the \textit{if} condition excludes several integers in the polyhedral space causing "holes" in the iteration space as shown in Figure~\subref{loop:c}. The excluded integers break the integrity of the space so that it no longer satisfies the definition of the convex set, thus the polyhedral model is not available for this particular circumstance. In this case, the true branch of the \textit{if} statement raises the problem however the false branch still satisfies the polyhedral model. Thus we can use the following equation to solve:
\begin{align*}
\scalebox{0.95}{$Count_{true\_branch} = Count_{loop\_total} - Count_{false\_branch} $}
\end{align*}
Because the counter of the outer loop and false branch both can be expressed by the polyhedral model, using either concrete value or parametric expression, so the count of the true branch is obtained. The generality of the polyhedral model makes it suitable for most common cases in real applications, however there are some cases that cannot be handled by the polyhedral model or even static analysis. For such circumstances, we provide users an option to annotate branches or the loops which Mira is not able to handle statically. 

\subsubsection{Annotation}
There are loop and branch cases that we are not able to process in a static way, such as conditionals involving loop index-unrelated variables or external function calls used for computing loop initial values or loop/branch conditions. Mira accepts user annotations to address such problems. We designed three types of annotation: an estimated percentage or a numerical value representing the proportion of iterations branch may take inside a loop or the number of iterations, which simplifies the loop/branch modeling; a variable used as initial value or condition to complete the polyhedral model; or a flag to indicate that a structure or a scope should be skipped. To annotate the code, users just need to put the information in a "\#pragma" directive in this format: \textit{\#pragma @Annotation {information}}. Mira processes the annotations during metric generation.

\begin{lstlisting}[caption={User annotation for \textit{if} statement}, frame=bt, belowcaptionskip=0.8em, label={list:6}]  
   for(i = 1; i <= 4; i++)
    for(j = a[i]; j <= a[i+6]; j++)
    {
     #pragma @Annotation \
            {lp_init:x,lp_cond:y}
     if(foo(i) > 10)
     {
      #pragma @Annotation {skip:yes} 
      statements;
     }    
    }
\end{lstlisting}

As the example shown in Listing~\ref{list:6}, the \textit{if} has a function call as a condition which causes a failure when Mira tries to generate the model fully automatically. To solve this problem, we specify an annotation in the pragma to provide the missing information and enable Mira to generate a complete model. In the given example, the whole branch scope will be skipped when generating metrics. Besides, we also annotate the initial value and condition of the inner loop using variable \textit{x} and \textit{y} because as a static tool Mira is not able to obtain values from those arrays. Mira will use the two variables to complete the polyhedral model; these variables will be treated as parameters expecting sample values from the user at model evaluation time.

\subsubsection{Functions}
Mira organizes the generated model in functions, which correspond to functions in the source code. In the generated model, the function header is modified for two reasons: flexibility and usability. Specifically, each user-defined function in the source code is modeled into a corresponding Python function with a different function signature, which only includes the arguments that are used \emph{by the model}. In addition, the generated model function has a slightly different name in order to avoid potential conflict due to different calling contexts or function overloading. For instance, the Python function with name \textit{foo\_2} represents the original C++ function \textit{foo}, but with a reduced number of arguments. In the body of the generated Python function, the original C++ statements are replaced with corresponding instruction counter metrics retrieved from binary. These data are stored in Python dictionaries and updated in the same order as the statements. Each function, when called, returns the aggregate counts within its scope. The advantage of this design is to provide the user the freedom to separate and obtain the overview of the particular functions with only minor changes to the model. 

Correct handling of function calls involves two aspects: locating the corresponding function and combining the metrics into the caller function. To combine the metrics, we designed a Python helper function \textit{handle\_function\_call}, which takes three arguments: caller metrics, callee metrics and loop iterations. It enables Mira to model the function call in the loop, which each metric of the callee should multiply the loop iterations. Mira retrieves the name of the callee function from the source AST node, and then generates a function call statement in Python and takes the return values that representing the metrics in the callee function. After that, Mira calls the \textit{handle\_function\_call} to combine metrics of the caller and the callee.

\subsubsection{Architecture Description File}
\begin{figure}
  \centering
    \includegraphics[width=0.45\textwidth]{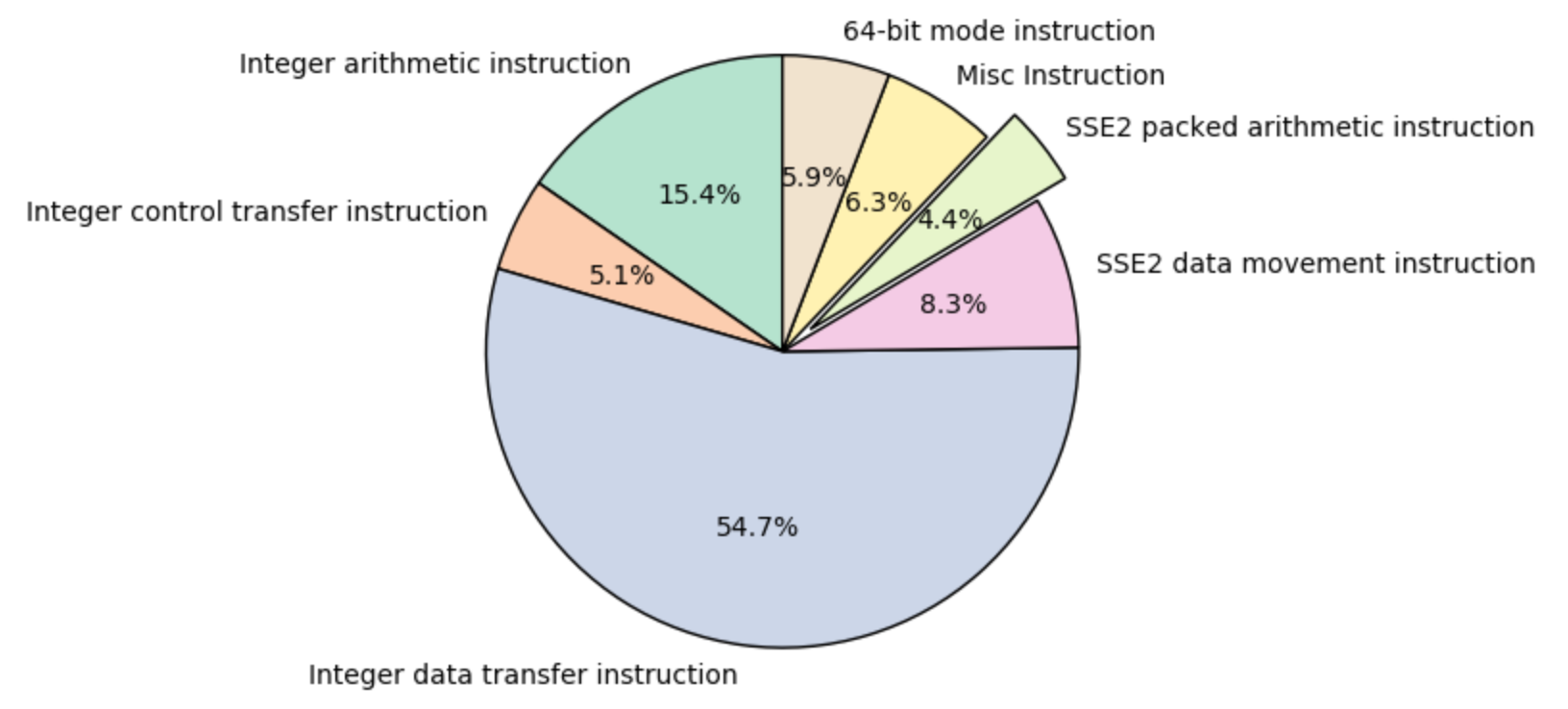}
  \caption{Instruction distribution of function cg\_solve}
  \label{fig:category}
\end{figure}

To enable the evaluation of the generated performance model in the context of specific architectural features, we provide an architecture description file where users define architecture-related parameters, such as number of CPU cores, cache line size, and vector length. Moreover, this user-customized file can be extended to include the information which does not exist in source or binary file to enable Mira to generate more predictions. For instance, we divided the x86 instruction set into 64 different categories in the description file, which Mira  uses to estimate the number of instructions in each category for each function in the source file. This representation strikes a balance between fine and coarse-grained approaches, providing category-based cumulative instruction counts at fine code granularity (down to statement-level), which enables developers to obtain better understanding of local behavior. Based on the metrics Mira generated in Table~\ref{tab:2}, Figure~\ref{fig:category} illustrates the distribution of categorized instructions from function \textit{cg\_solve} from the miniFE application~\cite{Mantevo}. The separated piece represents the SSE2 vector instructions which is the source of the floating-point instruction in this function.

\begin{table}
\centering
\caption{Categorized Instruction Counts of Function cg\_solve}
\label{tab:2}
\begin{tabular}{|c|l|}
\hline
Category                             & Count  \\ \hline
Integer arithmetic instruction       & 6.8E8  \\ \hline
Integer control transfer instruction & 2.26E8 \\ \hline
Integer data transfer instruction    & 2.42E9 \\ \hline
SSE2 data movement instruction       & 3.67E8 \\ \hline
SSE2 packed arithmetic instruction   & 1.93E8 \\ \hline
Misc Instruction                     & 2.77E8 \\ \hline
64-bit mode instruction              & 2.59E8 \\ \hline
\end{tabular}
\end{table}

\subsubsection{Generated Model}
We describe the model generated (output) by Mira with an example. In Figure~\ref{fig:model}, it shows the source code (input) and generated Python model separately. The source code (Figure~\subref*{model:a}) includes a class 
\textit{A} defining a member function \textit{foo} with two array variables as the parameters. The member function \textit{foo} is composed of a nested loop in which we annotate the upper bound of the inner loop with variable \textit{y}. In the \textit{main} function, it creates an instance of class \textit{A} and call function \textit{foo}. Figure~\subref*{model:b} shows part of the generated Python function \textit{foo} in which the new function name is replaced with the combination of its class name, original function name and the number of arguments in the original function definition. The body of the generated function \textit{A\_foo\_2} consists of the Python statements for keeping track of performance metrics. As we can see in the generated function, Mira uses the annotation variable \textit{y} to complete the polyhedral model and preserves \textit{y} as the argument. Similarly, the generated function \textit{main} is shown in Figure~\subref*{model:c}. It calls the \textit{A\_foo\_2} function and then updates its metrics by invoking \textit{handle\_function\_call}. The parameter \textit{y\_16} indicates that the function call associates the source code at line 16. At present, the value of \textit{y\_16} is specified by users during model evaluation. Different values can be supplied as function parameters in different function call contexts.

\begin{table}[t]
\centering
\caption{FPI Counts in STREAM benchmark}
\label{tab:3}
\begin{tabular}{|c|l|l|l|}
\hline
Array size / Tool & TAU     & Mira  &Error    \\ \hline\hline
2M          & 8.239E7 & 8.20E7 & 0.47\% \\ \hline
50M      & 4.108E9  & 4.100E9 & 0.19\% \\ \hline
100M      & 2.055E10 & 2.050E10 & 0.24\% \\ \hline
\end{tabular}
\end{table}

\begin{table}[t]
\centering
\caption{FPI Counts in DGEMM benchmark}
\label{tab:4}
\begin{tabular}{|c|l|l|l|}
\hline
Matrix size / Tool & TAU     & Mira  &Error    \\ \hline\hline
256          &1.013E9  & 1.0125E9 & 0.05\% \\ \hline
512      & 8.077E9  & 8.0769E9 & 0.0012\% \\ \hline
1024      & 6.452E10 & 6.4519E10 & 0.0015\% \\ \hline
\end{tabular}
\end{table}

\section{Evaluation}\label{sec:result}
In this section, we evaluate the correctness of the model derived by Mira with TAU in instrumentation mode. Two benchmarks are separately executed statically and dynamically on two different machines. While Mira counts all types of instructions, we focus on floating-point instructions (FPI) in this section because it is an important metric for HPC code analysis. The validation is performed by comparing the floating-point instruction counts produced by Mira with empirical instrumentation-based TAU/PAPI measurements. 

\subsection{Experiment environment}
We conducted the validation on two machines whose specifications are as follows.

\begin{itemize}
  \item \textbf{Arya} - Two Intel Xeon E5-2699v3 2.30GHz 18-core Haswell CPUs and 256GB of memory.
  \item \textbf{Frankenstein} - Two Intel Xeon E5620 2.40GHz 4-core Nehalem CPUs and 22GB of memory.
\end{itemize}

\subsection{Benchmarks}
Two benchmarks are chosen for validation, STREAM~\cite{mccalpin1995stream} and DGEMM~\cite{luszczek2006hpc}. 
STREAM is designed for the measurement of sustainable memory bandwidth and corresponded computation rate for simple vector kernels. DEGMM is a widely used benchmark for measuring the floating-point rate on a single CPU. It uses double-precision real matrix-matrix multiplication to calculate the floating-point rate. For both benchmarks, the non-OpenMP version is selected and executed serially with one thread.

\begin{table*}[t]
\centering
\caption{FPI Counts in miniFE}
\label{tab:5}
\begin{tabular}{|c|c|c|c|c|}
\hline
size     & Function / Tool         & TAU     & Mira    & Error   \\ \hline\hline
         & waxpby                  & 8.95E4  & 8.94E4  & 0.011\% \\ \cline{2-5} 
30x30x30 & matvec\_std::operator() & 1.54E6  & 1.52E6  & 1.3\%   \\ \cline{2-5} 
         & cg\_solve               & 1.966E8 & 1.925E8 & 2.09\%  \\ \hline
         & waxpby                  & 2.039E5 & 2.037E5 & 0.098\% \\ \cline{2-5} 
35x40x45 & matvec\_std::operator() & 3.57E6  & 3.46E6  & 3.08\%  \\ \cline{2-5} 
         & cg\_solve               & 7.621E8 & 7.386E8 & 3.08\%  \\ \hline
\end{tabular}
\end{table*}

\begin{figure*}
\centering

\subfloat[FP instruction counts in STREAM benchmark]{
	\includegraphics[width=0.40\textwidth]{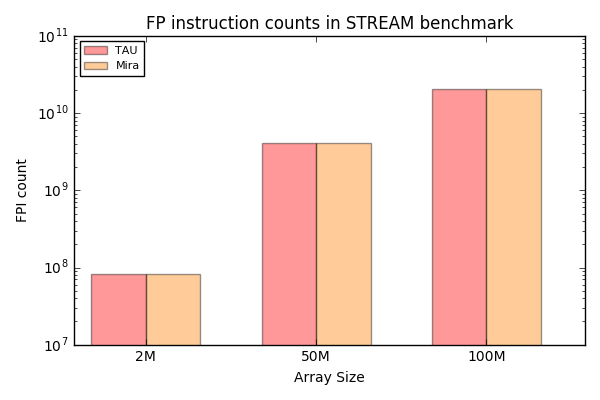}
	\label{val:a}
}
\subfloat[FP instruction counts in DGEMM benchmark]{
	\includegraphics[width=0.40\textwidth]{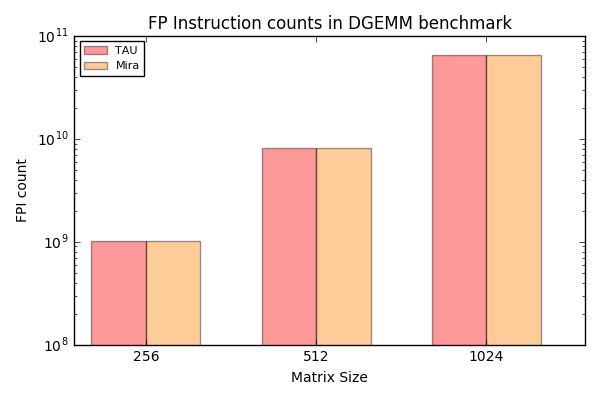}
	\label{val:b}
}	

\subfloat[FP instruction counts in miniFE]{
	\includegraphics[width=0.40\textwidth]{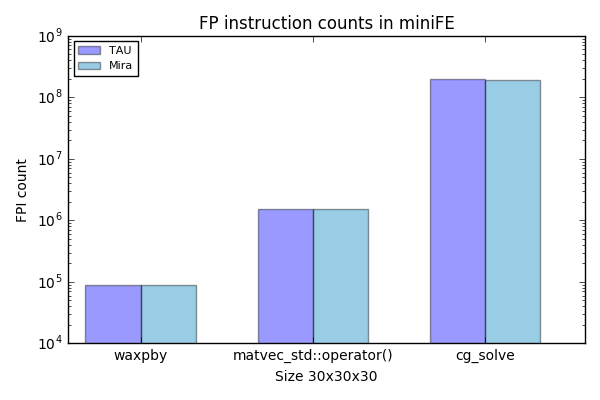}
	\label{val:c}
}
\subfloat[FP instruction counts in miniFE]{
	\includegraphics[width=0.40\textwidth]{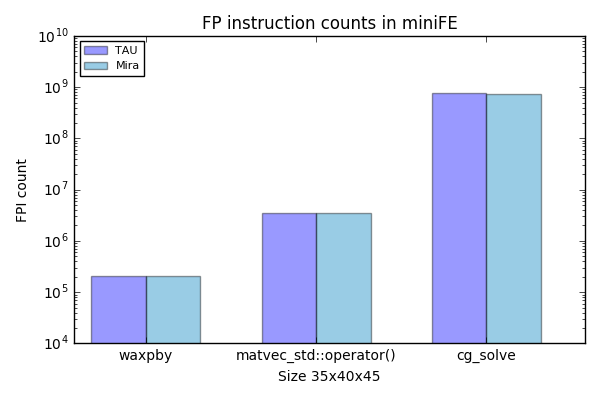}
	\label{val:d}
}
\caption{Validation of floating-point instruction counts.}
\label{fig:val}
\end{figure*}

\subsection{Mini Application}
In addition to the STREAM and DGEMM benchmarks, we also use the miniFE mini-application~\cite{Mantevo} to verify the result of Mira. MiniFE is composed of several finite-element kernels, including computation of element operators, assembly, sparse matrix-vector product, and vector operations. It assembles a sparse linear system and then solves it using a simple unpreconditioned conjugate-gradient algorithm. Unlike STREAM and DGEMM in which the \textit{main} function takes the majority part of the code, miniFE distributes the kernels by several functions, and associates each other by function calls which challenges the capability of Mira to handle a long chain of function calls.

\subsection{Results}
In this section, we present empirical validation results and illustrate the tradeoffs between static and dynamic methods for performance analysis and modeling. We also show a use case for the generated instruction metrics to compute an instruction-based arithmetic intensity derived metric, which can be used to identify loops that are good candidates for different types of optimizations (e.g., parallelization or memory-related tuning).

\subsubsection{Discussion}
Tables~\ref{tab:3}, \ref{tab:4} and \ref{tab:5} show floating-point instruction counts in two benchmarks and mini application separately. The metrics are gathered by evaluating the model generated by Mira, and comparing to the empirical results obtained through instrumentation-based measurement using TAU and PAPI. 

In Figure~\subref*{val:a}, the X axis is the size of the input array, and we choose 20 million, 50 million and 100 million, respectively. The logarithmic Y axis shows floating-point instruction counts. Similarly, in Figure~\subref*{val:b}, the X axis is for input size and the Y for FPI counts. Figure~\subref*{val:c} and Figure~\subref*{val:d} show FPI counts for three functions for the different problem sizes. We show details for the \textit{cg\_solve} function, which solves the sparse linear system, because it accounts for the bulk of the floating-point computations in this mini app. The function \textit{waxpby} and the operator overloading function \textit{matvec\_std::operator()} are in \textit{cg\_solve's} call tree and are invoked in the loop. 
Our results show that the floating-point instruction counts produced by Mira are close to the TAU measurements (of the PAPI\_FP\_INS values), with error of up to 3.08\%. The difference between static estimates and measured quantities increases with problem size, which means that there are discrepancies within some of the loops. This is not unexpected---static analysis cannot capture dynamic behavior with complete accuracy. The measured values capture samples based on all instructions, including those in external library function calls, which at present are not visible and hence not analyzed by Mira. For such scenarios, Mira can only track the function call statements that just contain several stack manipulation instructions while the content of the invoked function is skipped. In future we plan to provide different mechanisms for handling these cases, including limited binary analysis of the corresponding library supplemented by user annotations.

In addition to correctness, we compare the execution time of the static and empirical approaches. In empirical approaches, the experiment has to be repeated for different input values and in some cases multiple runs for each input value are required (e.g., when collecting performance hardware counter data). Instrumentation approaches can focus on specific code regions, but most sampling-based approaches collect information for all instructions, hence they potentially incur runtime and memory cost for collecting data on uninteresting instructions. By contrast, our model only needs to be generated once, and then can be evaluated (at low computational cost) for different user inputs and specific portions of the computation. Most important, the performance analysis by a parametric model can be used to achieve broad coverage without incurring the costs of many application executions.

Another challenge in dynamic approaches is the differences in hardware performance counters, including lack of availability of some types of measurements. For example, in modern Intel Haswell servers, there is no support for FLOP or FPI performance hardware counters. Hence, static performance analysis may be the only way to produce floating-point-based metrics in such cases.

\subsubsection{Prediction}
Next, we demonstrate how one can use the Mira-generated metrics to model the estimated instruction-based floating-point arithmetic intensity of the \textit{cg\_solve} function. The general definition of arithmetic intensity is the ratio of arithmetic operation to the memory traffic. With an appropriate setting in the architecture description file, we can enable Mira to generate various metrics. As the data shown in Table~\ref{tab:2}, Mira categorizes the instructions in \textit{cg\_solve} into seven categories. In the listed categories, "SSE2 packed arithmetic instruction" represents the packed and scalar double-precision floating-point instructions and "SSE2 data movement instruction" describes the movement of double-precision floating-point data between XMM registers and memory. Therefore the instruction-based floating-point arithmetic intensity of function \textit{cg\_solve} can be simply calculated as $ 1.93E8/3.67E8 = 0.53 $. This is a simple example to demonstrate the usage of our model. With sophisticated setting of architecture description file, Mira is able to perform more complicated prediction.

\section{Related Work}\label{sec:related}
There are two related tools that we are aware of designed for static performance modeling, PBound~\cite{PBound} and Kerncraft~\cite{hammer2015automatic}. PBound was designed by one of the authors of this paper (Norris) to estimate ``best case" or \emph{upper} performance bounds of C/C++ applications through static compiler analysis. It collects information and generates parametric expression for particular operations including memory access and floating-point operations, which is combined with user-provided architectural information to compute machine-specific performance estimates. However, it relies purely on source code analysis, and ignores the effects of compiler transformations (e.g., compiler optimization), frequently resulting in bound estimates that are not realistically achievable. 

Hammer et al. have created Kerncraft, a static performance modeling tool with concentration on memory hierarchy. Kerncraft characterizes performance and scaling loop behavior based on Roofline~\cite{williams2009roofline} or Execution-Cache-Memory (ECM)~\cite{hofmann2015execution} model. It uses YAML as the file format to describe low-level architecture and Intel Architecture Code Analyzer (IACA)~\cite{intel} to operate on binaries in order to gather loop-relevant information. However, the reliance on IACA limits the applicability of the tool so that the binary analysis is restricted by Intel architecture and compiler. 

Tools such as PDT\cite{Lindlan:2000:TFS:370049.370456} provide source-level
instrumentation, while MAQAO\cite{Djoudi2005} and Dyninst~\cite{Bernat:2011:AAB:2024569.2024572} use binary instrumentation for dynamic program analysis. Apollo~\cite{apollo} is a recent API-based dynamic analysis tool that provides a lightweight approach based on machine learning to select the best tuning parameter values while reducing the modeling cost by spreading it over multiple runs instead of constructing the model at runtime.

System simulators can also used for modeling, for example, the Structural Simulation Toolkit (SST)~\cite{sst}. However, as a system simulator, SST has a different focus---it simulates the whole system instead of single applications and it analyzes the interaction among architecture, programming model and communications system. Moreover, simulation is computationally expensive and limits the size and complexity of the applications that can be simulated. Compared with PBound, Kerncraft and Mira, SST is relatively heavyweight, complex, and focuses on hardware, which is more suitable for exploring architecture, rather than performance of the application. 

\section{Conclusion}\label{sec:conclusion}
In this paper, we present Mira, a framework for static performance modeling and analysis. We aim at designing a faster, accurate and flexible method for performance modeling as a supplement to existing tools in order to address problems that cannot solved by current tools. Our method focuses on floating-point operations and achieves good accuracy for benchmarks. These preliminary results suggest that this can be an effective method for performance analysis. 

While at present Mira can successfully analyze realistic application codes in many cases, much work remains to be done. In our future work, the first problem we eager to tackle is to appropriately handle more function-calling scenarios, especially those from system or third-party libraries. We will also consider combining dynamic analysis and introducing more performance metrics into the model to accommodate cases where control flow cannot be characterized accurately purely through static analysis. We also plan to extend Mira to enable characterization of shared-memory parallel programs.




\bibliographystyle{IEEEtran}
\bibliography{IEEEabrv,references}

\begin{thebibliography}{10}
\providecommand{\url}[1]{#1}
\csname url@samestyle\endcsname
\providecommand{\newblock}{\relax}
\providecommand{\bibinfo}[2]{#2}
\providecommand{\BIBentrySTDinterwordspacing}{\spaceskip=0pt\relax}
\providecommand{\BIBentryALTinterwordstretchfactor}{4}
\providecommand{\BIBentryALTinterwordspacing}{\spaceskip=\fontdimen2\font plus
\BIBentryALTinterwordstretchfactor\fontdimen3\font minus
  \fontdimen4\font\relax}
\providecommand{\BIBforeignlanguage}[2]{{%
\expandafter\ifx\csname l@#1\endcsname\relax
\typeout{** WARNING: IEEEtran.bst: No hyphenation pattern has been}%
\typeout{** loaded for the language `#1'. Using the pattern for}%
\typeout{** the default language instead.}%
\else
\language=\csname l@#1\endcsname
\fi
#2}}
\providecommand{\BIBdecl}{\relax}
\BIBdecl

\bibitem{PBound}
S.~H.~K. Narayanan, B.~Norris, and P.~D. Hovland, ``Generating performance
  bounds from source code,'' in \emph{Parallel Processing Workshops (ICPPW),
  2010 39th International Conference on}.\hskip 1em plus 0.5em minus
  0.4em\relax IEEE, 2010, pp. 197--206.

\bibitem{williams2009roofline}
S.~Williams, A.~Waterman, and D.~Patterson, ``Roofline: an insightful visual
  performance model for multicore architectures,'' \emph{Communications of the
  ACM}, vol.~52, no.~4, pp. 65--76, 2009.

\bibitem{rose}
D.~Quinlan, ``Rose homepage,'' http://rosecompiler.org.

\bibitem{pouchet.11.popl}
L.-N. Pouchet, U.~Bondhugula, C.~Bastoul, A.~Cohen, J.~Ramanujam,
  P.~Sadayappan, and N.~Vasilache, ``Loop transformations: Convexity, pruning
  and optimization,'' in \emph{38th ACM SIGACT-SIGPLAN Symposium on Principles
  of Programming Languages (POPL'11)}.\hskip 1em plus 0.5em minus 0.4em\relax
  Austin, TX: ACM Press, Jan. 2011, pp. 549--562.

\bibitem{Bas04b}
C.~Bastoul, ``Code generation in the polyhedral model is easier than you
  think,'' in \emph{PACT'13 IEEE International Conference on Parallel
  Architecture and Compilation Techniques}, Juan-les-Pins, France, September
  2004, pp. 7--16.

\bibitem{griebl2004automatic}
M.~Griebl, ``Automatic parallelization of loop programs for distributed memory
  architectures,'' 2004.

\bibitem{bondhugula2008practical}
U.~Bondhugula, A.~Hartono, J.~Ramanujam, and P.~Sadayappan, ``A practical
  automatic polyhedral parallelizer and locality optimizer,'' in \emph{ACM
  SIGPLAN Notices}, vol.~43, no.~6.\hskip 1em plus 0.5em minus 0.4em\relax ACM,
  2008, pp. 101--113.

\bibitem{grosser2012polly}
T.~Grosser, A.~Groesslinger, and C.~Lengauer, ``Polly—performing polyhedral
  optimizations on a low-level intermediate representation,'' \emph{Parallel
  Processing Letters}, vol.~22, no.~04, p. 1250010, 2012.

\bibitem{mucci1999papi}
P.~J. Mucci, S.~Browne, C.~Deane, and G.~Ho, ``Papi: A portable interface to
  hardware performance counters,'' in \emph{Proceedings of the department of
  defense HPCMP users group conference}, 1999, pp. 7--10.

\bibitem{shende2006tau}
S.~S. Shende and A.~D. Malony, ``The tau parallel performance system,''
  \emph{International Journal of High Performance Computing Applications},
  vol.~20, no.~2, pp. 287--311, 2006.

\bibitem{adhianto2010hpctoolkit}
L.~Adhianto, S.~Banerjee, M.~Fagan, M.~Krentel, G.~Marin, J.~Mellor-Crummey,
  and N.~R. Tallent, ``Hpctoolkit: Tools for performance analysis of optimized
  parallel programs,'' \emph{Concurrency and Computation: Practice and
  Experience}, vol.~22, no.~6, pp. 685--701, 2010.

\bibitem{geimer2010scalasca}
M.~Geimer, F.~Wolf, B.~J. Wylie, E.~{\'A}brah{\'a}m, D.~Becker, and B.~Mohr,
  ``The scalasca performance toolset architecture,'' \emph{Concurrency and
  Computation: Practice and Experience}, vol.~22, no.~6, pp. 702--719, 2010.

\bibitem{marin2014miami}
G.~Marin, J.~Dongarra, and D.~Terpstra, ``Miami: A framework for application
  performance diagnosis,'' in \emph{Performance Analysis of Systems and
  Software (ISPASS), 2014 IEEE International Symposium on}.\hskip 1em plus
  0.5em minus 0.4em\relax IEEE, 2014, pp. 158--168.

\bibitem{graham1982gprof}
S.~L. Graham, P.~B. Kessler, and M.~K. Mckusick, ``Gprof: A call graph
  execution profiler,'' in \emph{ACM Sigplan Notices}, vol.~17, no.~6.\hskip
  1em plus 0.5em minus 0.4em\relax ACM, 1982, pp. 120--126.

\bibitem{pakin2013hardware}
S.~Pakin and P.~McCormick, ``Hardware-independent application
  characterization,'' in \emph{Workload Characterization (IISWC), 2013 IEEE
  International Symposium on}.\hskip 1em plus 0.5em minus 0.4em\relax IEEE,
  2013, pp. 111--112.

\bibitem{bastoul2003putting}
C.~Bastoul, A.~Cohen, S.~Girbal, S.~Sharma, and O.~Temam, ``Putting polyhedral
  loop transformations to work,'' in \emph{International Workshop on Languages
  and Compilers for Parallel Computing}.\hskip 1em plus 0.5em minus 0.4em\relax
  Springer, 2003, pp. 209--225.

\bibitem{Mantevo}
M.~A. Heroux, D.~W. Doer<DF>er, P.~S. Crozier, J.~M. Willenbring, H.~C.
  Edwards, A.~Williams, M.~Rajan, E.~R. Keiter, H.~K. Thornquist, and R.~W.
  Numrich, ``Improving performance via mini-applications,'' Sandia National
  Laboratories, Tech. Rep. SAND2009-5574, Sept. 2009.

\bibitem{mccalpin1995stream}
J.~D. McCalpin, ``A survey of memory bandwidth and machine balance in current
  high performance computers,'' \emph{IEEE TCCA Newsletter}, vol.~19, p.~25,
  1995.

\bibitem{luszczek2006hpc}
P.~R. Luszczek, D.~H. Bailey, J.~J. Dongarra, J.~Kepner, R.~F. Lucas,
  R.~Rabenseifner, and D.~Takahashi, ``The hpc challenge (hpcc) benchmark
  suite,'' in \emph{Proceedings of the 2006 ACM/IEEE conference on
  Supercomputing}.\hskip 1em plus 0.5em minus 0.4em\relax Citeseer, 2006, p.
  213.

\bibitem{hammer2015automatic}
J.~Hammer, G.~Hager, J.~Eitzinger, and G.~Wellein, ``Automatic loop kernel
  analysis and performance modeling with kerncraft,'' in \emph{Proceedings of
  the 6th International Workshop on Performance Modeling, Benchmarking, and
  Simulation of High Performance Computing Systems}.\hskip 1em plus 0.5em minus
  0.4em\relax ACM, 2015, p.~4.

\bibitem{hofmann2015execution}
J.~Hofmann, J.~Eitzinger, and D.~Fey, ``Execution-cache-memory performance
  model: Introduction and validation,'' \emph{arXiv preprint arXiv:1509.03118},
  2015.

\bibitem{intel}
Intel, ``Intel architecture code analyzer homepage,''
  https://software.intel.com/en-us/articles/intel-architecture-code-analyzer.

\bibitem{Lindlan:2000:TFS:370049.370456}
\BIBentryALTinterwordspacing
K.~A. Lindlan, J.~Cuny, A.~D. Malony, S.~Shende, F.~Juelich, R.~Rivenburgh,
  C.~Rasmussen, and B.~Mohr, ``A tool framework for static and dynamic analysis
  of object-oriented software with templates,'' in \emph{Proceedings of the
  2000 ACM/IEEE Conference on Supercomputing}.\hskip 1em plus 0.5em minus
  0.4em\relax IEEE Computer Society, 2000. [Online]. Available:
  \url{http://dl.acm.org/citation.cfm?id=370049.370456}
\BIBentrySTDinterwordspacing

\bibitem{Djoudi2005}
L.~Djoudi and D.~Barthou, ``{Maqao}: Modular assembler quality analyzer and
  optimizer for {I}tanium 2,'' \emph{Workshop on EPIC architectures and
  compiler technology}, 2005.

\bibitem{Bernat:2011:AAB:2024569.2024572}
\BIBentryALTinterwordspacing
A.~R. Bernat and B.~P. Miller, ``Anywhere, any-time binary instrumentation,''
  in \emph{Proc. of the 10th ACM SIGPLAN-SIGSOFT Workshop on Program Analysis
  for Software Tools}, ser. PASTE '11.\hskip 1em plus 0.5em minus 0.4em\relax
  ACM, 2011, pp. 9--16. [Online]. Available:
  \url{http://doi.acm.org/10.1145/2024569.2024572}
\BIBentrySTDinterwordspacing

\bibitem{apollo}
D.~Beckingsale, O.~Pearce, I.~Laguna, and T.~Gamblin, ``Apollo: Reusable models
  for fast, dynamic tuning of input-dependent code,'' in \emph{The 31th IEEE
  International Parallel and Distributed Processing Symposium}, 2017.

\bibitem{sst}
A.~Rodrigues, K.~S. Hemmert, B.~W. Barrett, C.~Kersey, R.~Oldfield, M.~Weston,
  R.~Riesen, J.~Cook, P.~Rosenfeld, E.~Cooper-Balis, and B.~Jacob, ``The
  structural simulation toolkit,'' \emph{SIGMETRICS Perform. Eval. Rev.},
  vol.~38, no.~4, pp. 37--42, March 2011.

\end{thebibliography}

\end{document}